\documentclass[aps,pra,twocolumn,10pt,superscriptaddress]{revtex4-1}
\usepackage{graphicx,amsmath,amssymb}
\usepackage[utf8x]{inputenc}
\usepackage{url}
\usepackage{float}
\usepackage{bm} 
\usepackage{enumitem}
\usepackage{hyperref}
\usepackage{threeparttable}

\begin{document}

\title{Global polarizability matrix method for efficient modeling of light scattering by dense ensembles of non-spherical particles in stratified media}

\author{Maxime Bertrand}
\affiliation{LP2N, CNRS, Institut d'Optique Graduate School, Univ. Bordeaux, 33400 Talence, France}

\author{Alexis Devilez}
\affiliation{LP2N, CNRS, Institut d'Optique Graduate School, Univ. Bordeaux, 33400 Talence, France}

\author{Jean-Paul Hugonin}
\affiliation{Laboratoire Charles Fabry, CNRS, Institut d'Optique Graduate School, Universit\'{e} Paris-Saclay, 91127 Palaiseau, France}

\author{Philippe Lalanne}
\email{philippe.lalanne@institutoptique.fr}
\affiliation{LP2N, CNRS, Institut d'Optique Graduate School, Univ. Bordeaux, 33400 Talence, France}

\author{Kevin Vynck}
\email{kevin.vynck@institutoptique.fr}
\affiliation{LP2N, CNRS, Institut d'Optique Graduate School, Univ. Bordeaux, 33400 Talence, France}

\date{\today}

\begin{abstract}
We introduce a numerical method that enables efficient modelling of light scattering by large, disordered ensembles of non-spherical particles incorporated in stratified media, including when the particles are in close vicinity to each other, to planar interfaces and/or to localized light sources. The method consists in finding a small set of fictitious polarizable elements -- or numerical dipoles -- that quantitatively reproduces the field scattered by an individual particle for any excitation and at an arbitrary distance from the particle surface. The set of numerical dipoles is described by a global polarizability matrix that is determined numerically by solving an inverse problem relying on fullwave simulations. The latter are classical and may be performed with any Maxwell's equations solver. Spatial non-locality is an important feature of the numerical dipoles set, providing additional degrees of freedom compared to classical coupled dipoles to reconstruct complex scattered fields. Once the polarizability matrix describing scattering by an individual particle is determined, the multiple scattering problem by ensembles of such particles in stratified media can be solved using a Green tensor formalism and only few numerical dipoles, thereby with a low physical memory usage, even for dense systems in close vicinity to interfaces. The performance of the method is studied with the example of large high-aspect-ratio high-index dielectric cylinders. The method is easy to implement and may offer new possibilities for the study of complex nanostructured surfaces, which are becoming widespread in emerging photonic technologies.
\end{abstract}

\maketitle

\section{Introduction}

Theoretical and numerical modelling of light scattering by particles is a long-standing topic that experienced many advances and outcomes especially in the context of atmospheric optics~\cite{hulst1981light, bohren2008absorption}. The research field took a new twist in the past decades with the advent of nanophotonics and plasmonics, in which nanostructures are designed to control the emission, propagation and confinement of light~\cite{joannopoulos2008molding, agio2013optical}. Nowadays, a myriad of electromagnetic nanoparticles, possibly mixing dielectric and metallic materials and having a complex shape, can be synthesized on large quantities by colloidal means~\cite{xia2005shape, sau2010nonspherical}. These nanoparticles exhibit a rich panel of optical features such as strong scattering and absorption efficiencies that can be tuned on the whole visible and near-infrared ranges~\cite{oldenburg1998nanoengineering} or a controlled directivity of the emitted or scattered light~\cite{mirin2009light, fu2013directional}. The self-assembly of such nanoparticles in stratified media can further enrich their optical properties thanks to the interaction of the nanoparticles with planar interfaces and between themselves~\cite{castanie2012absorption, miroshnichenko2015substrate, leseur2016high}. One may obtain, for instance, a frequency-selective wide-angle absorption or emission thanks to hybrid nanoparticle-stack modes~\cite{akselrod2015large, faggiani2015quenching} or a controlled light extraction from light-emitting materials by engineering the nanoparticles and their relative position in a disordered array~\cite{gomard2016photon, jouanin2016designer}. The self-assembly of nanoresonators with electric and magnetic dipole resonances also enables the realization of (meta)materials with strongly dispersive effective permittivities and permeabilities~\cite{muhlig2011optical, baron2016self}, designed to mold wave propagation. Predicting quantitatively the optical response of complex nanostructures, consisting of disordered ensembles of non-spherical particles incorporated in stratified media, requires being able to model both coherent phenomena occurring at the level of an individual particle (nano-scale) and at the level of an ensemble of interacting particles (meso-scale).

Various numerical methods for solving Maxwell's equations in complex nanostructures have been developed over the years~\cite{gallinet2015numerical}. Owing to their great flexibility, methods relying on a discretization of the entire computational domain (e.g., finite-differences, finite-elements) are powerful tools for studying small systems but are not adapted to the modelling of spatially-extended systems made of many particles. In such cases, it is preferable to rely on the high degree of analyticity of dyadic Green functions describing propagation in uniform or stratified media to limit the discretization to the volume or even just the surface of the particles. Popular approaches in this category include the discrete dipole approximation~\cite{draine1994discrete}, the boundary element method~\cite{buffa2003galerkin} and the surface integral equation method~\cite{kern2009surface}. Because these methods rely on a discretization of the physical system, they are expected to predict the field with high accuracy -- remarkable results have been obtained for instance in Ref.~\cite{solis2014toward}.

Largely developed by several groups in the 1990s~\cite{leviatan1987analysis, hafner1990generalized, zolla1994electromagnetic, wriedt2000acoustic, eremin2007mathematical}, methods known as the generalized multipole technique, the multiple multipole method, the discrete sources method or the method of fictitious sources, to cite only a few, rely on a markedly different concept, namely that a scattering object may be represented by a small set of elementary multipole sources distributed in space. The amplitudes of the elementary sources, which are unrelated to the particle physical parameters, are determined by least-square minimization to match the boundary conditions on the surface of the particle and outgoing wave conditions. These methods lower the computational cost for solving scattering problems by yielding approximate solutions of the boundary-value problem~\cite{tikhonov1963solution, ivanov1966approximate}, which are however expected to converge to the exact solution upon a proper choice of the source basis. The methods have been successfully applied to describe diffraction and scattering by gratings and non-spherical particles~\cite{zolla1996method, moreno2002multiple, wriedt2018generalized}. However, because the numerical inverse problem to determine the source amplitudes should be solved for all boundaries at once and for a specific incident field, they are not ideally suited to multiple scattering by large disordered ensembles of particles.

Probably the most classical method for multiple scattering problems by large ensembles of particles is the so-called T-matrix method (where T stands for ``transition''). Originally introduced by Waterman~\cite{waterman1965matrix} and largely improved afterwards~\cite{mishchenko1996t}, the method relies on the possibility to expand the incident and scattered fields for an individual particle on a complete and orthonormal basis of vector spherical harmonics (VSHs). The T-matrix formally relates the electric and magnetic multipolar coefficients of the incident and scattered fields. Like the polarizability, it is an intrinsic quantity describing the particle at a given frequency. The T-matrix of a particle can in principle be computed with any Maxwell's equations solver~\cite{doicu1999calculation, mackowski2002discrete, fruhnert2017computing}. Once known, multiple scattering problems by particle ensembles can be solved by a proper description of translation operations between particles~\cite{kristensson1980electromagnetic, xu1996calculation, mishchenko2006multiple, stout2008recursive, egel2016efficient}. Since few multipolar orders are generally sufficient for most wavelength-scale particles, the problem becomes computationally very efficient.

The major limitation of the T-matrix method is its incapability to deal with elongated particles placed in the near-field of each other or of an interface. The reason is that the VSH decomposition of the field is formally valid only in a uniform background beyond the smallest sphere that circumscribes the entire particle. Important efforts have been dedicated to this basic problem. A first proposed solution to reconstruct the near field around a particle is to expand it into regular and radiating VSHs and then solve a boundary-value problem on the near-field domain, between the particle surface and the circumscribing sphere~\cite{bringi1980surface, doicu2010near}. The use of multipolar sources distributed throughout the particle volume and not restricted to the center of the circumscribing sphere improves the near-field computation for high-aspect-ratio spheroids~\cite{forestiere2011near}. Unfortunately, the excitation cannot be located in the near-field region, thereby excluding the possibility to simulate particles placed in the near-field of each other or near an interface. Besides, one may anticipate difficulties in solving scattering problems with particles of complex shapes (e.g., having sharp corners) and composition (e.g., mixing metallic and dielectric materials). A second proposed solution is to perform a planewave expansion of the radiating VSHs on a plane in the vicinity of the particle~\cite{egel2016light,egel2017extending}. With a proper choice of truncations for the planewave and VSH expansions and a proper modification of the translation operator, this enables solving multiple scattering problems as long as one can define a non-intersecting plane separating each pair of particles~\cite{theobald2017plane}. Due to multiple transformations between spherical waves and planewaves, the method may become computationally heavy for large ensembles of particles, nevertheless it has demonstrated remarkable accuracy on dense clusters of elongated nanoparticles~\cite{theobald2017plane}.

In this article, we introduce a numerical method, named global polarizability matrix (GPM) method, for light scattering by arbitrary (non-spherical) particles that remains highly accurate in the near-field region for both the excitation and observation points, that reduces the computational efforts and that is easy to implement. Inspired by numerical methods and near-field reconstruction techniques relying on distributed elementary sources~\cite{wriedt2000acoustic, beghou2009synthesis, chardon2013sampling}, the method consists in finding a small set of fictitious polarizable elements that can reproduce the field scattered by an individual particle for an arbitrary (near or far-field) excitation beyond a virtual surface of any shape and at any distance from the particle surface. This virtual surface thus replaces the circumscribing sphere of the T-matrix method. The set of ``numerical dipoles'' is described by a global polarizability matrix that is spatially non-local -- the moments induced at a dipole position depends on the background fields at all dipole positions. Numerical dipoles are thus very different from physical dipoles that would interact via electromagnetic wave propagation. Compared to physical dipoles that are described by local polarizabilities, like in the discrete dipole approximation~\cite{draine1994discrete}, the spatial non-locality implemented with numerical dipoles provides additional degrees of freedom that are expected to provide highly-accurate field reconstructions, even for moderate numbers of dipoles. The global polarizability matrix is determined numerically by solving an inverse problem relying on fullwave simulation data, which may be obtained from any Maxwell's equations solver. Obtaining the relevant dataset is the computationally-heavier step but, once done, the multiple scattering problem by a large ensemble of such particles in a stratified medium can be solved efficiently with a Green tensor formalism, even in cases of particles interacting in the near field of each other, of planar interfaces or of localized light sources.

The original concept of using distributed sources to reproduce the field scattered by an arbitrary particle for an arbitrary excitation, which is the core of the GPM method, has been developed by Hugonin several years ago. The method has remained unpublished but has been successfully used already to study cooperative absorption effects in nanoparticle arrays~\cite{langlais2014cooperative, hugonin2017photovoltaics} and light extraction by disordered ensembles of metallic nanoparticles in optical stacks~\cite{jouanin2016designer}. In these works, the polarizability matrix of the numerical dipoles set was determined to match the VSHs radiated by the individual scatterers for an arbitrary incident field. Like in the T-matrix method, the scattered field could thus only be predicted beyond the minimum circumscribing sphere. The major extension brought by the present work is to enable the modelling of non-spherical particles in close vicinity to each other, to planar interfaces and to localized light sources. This possibility stems from a genuine improvement of the modelling of the particle near-field with numerical dipoles.

The remainder of this article is structured as follows. Section~\ref{numericalmethod} introduces the GPM method for scattering by individual complex particles. The global polarizability matrix is defined and the numerical approach to compute it is described. The method is illustrated with the example of a high-aspect-ratio high-index dielectric cylinder. We compute the global polarizability matrix and investigate the convergence of the method with increasing number of numerical dipoles. We finally demonstrate the validity of the method to reproduce the scattered field around the particle for both near and far-field excitations via comparisons with reference data. Section~\ref{multiplescattering} is then concerned with the use of the GPM method to simulate light scattering by ensembles of particles in stratified media, possibly interacting in the near-field of each other or with interfaces. The convergence of scattering quantities (diagram and cross-section) is studied as a function of distance between the objects. We eventually report on a simulation on a dense ensemble of 16 resonant cylinders incorporated in a metallo-dielectric stack, yielding very good agreement with reference calculations for surface-to-surface distances of about $\lambda/(10n_\text{b})$, with $n_\text{b}$ the refractive index of the medium surrounding the cylinders. We conclude in Sec.~\ref{conclusion} with some perspectives and ideas for future studies. Appendix~\ref{App:Tmatrix} reminds the well-known concept of transition operator for light scattering to highlight some similarities and differences with the GPM method.

\section{Global polarizability matrix method}
\label{numericalmethod}

\subsection{Concept and formalism}

To deal with the problem of light scattering by ensembles of particles including when they are in close vicinity to each other, to planar interfaces or to localized light sources, one needs to be able to predict the near field surrounding a particle for an arbitrary near or far-field excitation. It is well comprehended that the rapid spatial variations of an optical near field around an object may be reproduced by distributing few multipole sources optimally-distributed throughout a volume. This concept is exploited since many years in various electromagnetic methods based on elementary sources~\cite{wriedt2000acoustic}, which have demonstrated excellent performances for direct light scattering problems on individual objects. Such near-field reconstruction has also been investigated in the area of microwave components. It was shown that the near field radiated from power electronic devices can be reproduced with a small set of electric and magnetic dipoles, whose amplitudes were determined by solving an inverse problem based on near-field measurements~\cite{beghou2009synthesis}. Related works are found, for instance, in acoustics, where loudspeakers are optimized to approximately match a target sound field over a region of interest~\cite{chardon2013sampling}. In all of these works, the elements are \textit{current sources}, producing either a total field in the case of emitting objects or a scattered field for a given incident field in the case of scattering objects.

\begin{figure}
   \centering
   \includegraphics[width=0.9\columnwidth]{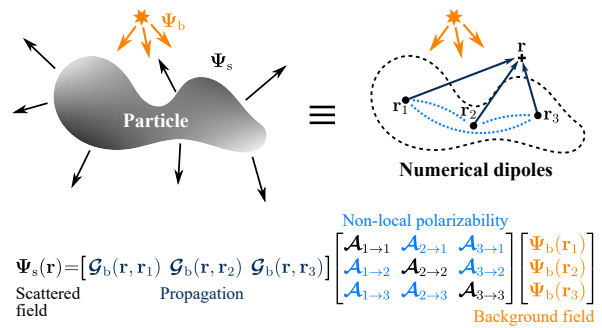}
   \caption{Illustration of the method. Light scattering by an arbitrary particle (left) is modelled by a small set of polarizable elements (right) capable of reproducing accurately the field $\bm{\Psi}_\text{s}= \left[\mathbf{E}_\text{s} ; \mathbf{H}_\text{s} \right]$ scattered by the particle everywhere around it, including in the near-field region, for an arbitrary background field $\bm{\Psi}_\text{b}= \left[\mathbf{E}_\text{b} ; \mathbf{H}_\text{b} \right]$. The numerical dipoles are described by a global polarizability tensor, that is spatially non-local because every induced moment depends not only on the incident field at the dipole position but also on the incident field at the positions of the other dipoles. The spatial non-locality is sketched here as dotted blue lines and is formally described by the off-diagonal elements of the matrix, $\bm{\mathcal{A}}_{i \rightarrow j}$, with $i \neq j$. The scattering problem formulation of Eq.~(\ref{eq:scattering-numerical-dipoles}) is repeated here for $N_\text{d}=3$.}
    \label{fig1}
\end{figure}

The method introduced here exploits the concept of near-field reconstruction with distributed sources. It consists in finding a set of \textit{polarizable} electric and magnetic dipoles capable of predicting the near field scattered by a particle for \textit{any} near or far-field excitation, see Fig.~\ref{fig1}.

Consider a monochromatic electromagnetic wave $\bm{\Psi}_\text{b} = \left[\mathbf{E}_\text{b} ; \mathbf{H}_\text{b} \right]$ at frequency $\omega = k_0 c$, with $k_0$ the wave number in vacuum and $c$ the speed of light, impinging on a particle with relative permittivity $\bm{\epsilon} (\mathbf{r}) = \epsilon_\text{b} \mathbf{I} + \bm{\Delta\epsilon}(\mathbf{r})$, where $\mathbf{r}$ is the position in Cartesian coordinates, $\epsilon_\text{b}$ is the permittivity of the background and $\bm{\Delta\epsilon}$ is the permittivity variation due to the particle. We formulate the electromagnetic scattering problem by posing that the scattered electromagnetic field $\bm{\Psi}_\text{s} = \left[\mathbf{E}_\text{s} ; \mathbf{H}_\text{s} \right]$ outside the particle can be accurately represented by a set of $N_\text{d}$ electric and magnetic dipoles as
\begin{equation}\label{eq:scattering-numerical-dipoles}
\bm{\Psi}_\text{s} (\mathbf{r}) = \sum_{i,j=1}^{N_\text{d}} \bm{\mathcal{G}}_\text{b} (\mathbf{r},\mathbf{r}_j) \bm{\mathcal{A}}(\mathbf{r}_j,\mathbf{r}_i) \bm{\Psi}_\text{b} (\mathbf{r}_i).
\end{equation}
In this expression, $\bm{\mathcal{G}}_\text{b}$ is the electromagnetic dyadic Green function
\begin{equation}\label{eq:Green-dyadic}
\bm{\mathcal{G}}_\text{b} =
\begin{bmatrix}
\mathbf{G}_\text{b}^\text{ee} & \mathbf{G}_\text{b}^\text{em}\\ 
\mathbf{G}_\text{b}^\text{me} & \mathbf{G}_\text{b}^\text{mm}
\end{bmatrix},
\end{equation}
where $\mathbf{G}_\text{b}^{\gamma \delta}$ is the dyadic Green function describing the electric and magnetic fields ($\delta=e$ or $m$) produced by electric and magnetic dipoles ($\gamma=e$ or $m$) in a uniform medium of refractive index $n_\text{b}=\sqrt{\epsilon_\text{b}}$~\cite{tai1994dyadic}.

Equation~(\ref{eq:scattering-numerical-dipoles}) can be understood as follows. A background electromagnetic field at the dipole position $\mathbf{r}_i$ yields electric and magnetic dipole moments at the dipole positions $\mathbf{r}_j$ via the polarizability tensor $\bm{\mathcal{A}}$. As illustrated in Fig.~\ref{fig1}, for $N_\text{d}=3$ for instance, the dipole moment induced in dipole $1$ would be $\mathbf{d}_1 = \bm{\mathcal{A}}(\mathbf{r}_1,\mathbf{r}_1) \bm{\Psi}_\text{b} (\mathbf{r}_1) + \bm{\mathcal{A}}(\mathbf{r}_1,\mathbf{r}_2) \bm{\Psi}_\text{b} (\mathbf{r}_2) + \bm{\mathcal{A}}(\mathbf{r}_1,\mathbf{r}_3) \bm{\Psi}_\text{b} (\mathbf{r}_3)$, and similarly for the other dipoles. The induced dipole moments at $\mathbf{r}_j$ then generate an electromagnetic field at point $\mathbf{r}$ via the Green tensors. The scattered field is straightforwardly recovered by performing a sum over all dipoles $i$ (for the background field) and $j$ (for the radiating dipoles). For the example of Fig.~\ref{fig1}, the scattered field would then be $\bm{\Psi}_\text{s} (\mathbf{r}) = \bm{\mathcal{G}}_\text{b} (\mathbf{r},\mathbf{r}_1) \mathbf{d}_1 + \bm{\mathcal{G}}_\text{b} (\mathbf{r},\mathbf{r}_2) \mathbf{d}_2 + \bm{\mathcal{G}}_\text{b} (\mathbf{r},\mathbf{r}_3) \mathbf{d}_3$.

The key ingredient in Eq.~(\ref{eq:scattering-numerical-dipoles}) is evidently the polarizability tensor $\bm{\mathcal{A}}$. It is essential to remark that $\bm{\mathcal{A}}$ is a property of the ensemble of dipoles and is spatially non-local (as sketched by the dashed blue lines in Fig.~\ref{fig1}). As we will show below, we exploit this spatial non-locality to provide additional degrees of freedom for field prediction compared to physical dipoles, which are described individually by \textit{local} polarizabilities and interact via classical electromagnetic wave propagation~\cite{draine1994discrete}. We call our elements ``numerical dipoles'' precisely to avoid a possible confusion with physical dipoles. It is also worth noticing that Eq.~(\ref{eq:scattering-numerical-dipoles}) closely ressembles the classical definition of the transition operator, which is also a spatially non-local quantity. As discussed in App.~\ref{App:Tmatrix}, unlike the transition operator, the global polarizability tensor is however not explicitly related to the actual shape and composition of the particle. It is a purely mathematical object introduced to solve the scattering problem efficiently.

The major question to address is how to find a polarizability tensor $\bm{\mathcal{A}}$ that yields accurate predictions of the scattered near and far-field for any near or far-field excitation. We propose to achieve this by building a dataset of \textit{independent} scattering solutions obtained from rigorous fullwave simulations using an external yet arbitrary Maxwell's equations solver (e.g. finite-elements), and consecutively by computing a global polarizability matrix that can reproduce the set of precomputed solutions by numerical inversion of the problem~\cite{vogel2002computational}. The set of precomputed solutions will hereafter be called the ``learning set''. The computed polarizability matrix is uniquely defined from the whole dataset. When using a sufficiently large number of dipoles, it is expected to reconstruct accurately the scattered field for all of the precomputed solutions. Because an accurate description of the field on a closed surface circumscribing the particle leads to an accurate description of it everywhere beyond this surface~\cite{balanis2016antenna}, learning may be limited to the scattered field on a closed surface.

Reproducing the scattered field for excitations that belong to the learning set is however not sufficient for our purpose. Indeed, to describe multiple scattering, one needs to \textit{predict} the scattered field for \textit{unknown} excitations. This is called ``generalization''. Because background field variations can be reproduced by summing several background fields, one expects that the use of a sufficiently large number of independent sources for learning can lead to accurate predictions of the field for any excitation. In the following subsections, we will provide evidence that these expectations are perfectly sound.

\subsection{Numerical implementation}

The GPM method, from the fullwave simulations to the field prediction, can be decomposed in six main steps.
\begin{itemize}[leftmargin=*]
\item \textbf{Step 1}: Using a rigorous Maxwell's equations solver, we compute the field scattered by a particle of interest for $N_\text{s}$ independent excitations at frequency $\omega$. The excitations may be planewaves with varying incident angles and polarizations or dipole sources (electric or magnetic) with varying dipole orientations and positions in space. The dataset contains the background and scattered electric and magnetic fields, $\bm{\Psi}_\text{b}$ and $\bm{\Psi}_\text{s}$, at many points around the particle. The number of points naturally depends on the method and discretization used. Considering many points at this step leaves more flexibility afterwards in the choice of the numerical dipoles spatial arrangement and of the learning surface. It is the only time-consuming task, but it has to be performed only once for a given particle at a given frequency.

\item \textbf{Step 2}: The closed surface on which the field should be accurately reproduced (hereafter called the ``learning surface'') is defined. This surface may be arbitrarily close to the particle boundaries and have \textit{a priori} any shape. It plays the role of the smallest circumscribing sphere of the T-matrix method. For each of the $N_\text{s}$ excitations, the six components of the electric and magnetic scattered fields $\bm{\Psi}_\text{s}$ are evaluated on $N_\text{l}$ learning points distributed regularly on the surface, resulting in a matrix $\bm{\mathsf{P}}_\text{s}^\text{true}$ of size $6 N_\text{l} \times N_\text{s}$.

\item \textbf{Step 3}: The position of the $N_\text{d}$ numerical dipoles in the volume of the particle is defined. The six components of the electric and magnetic background fields $\bm{\Psi}_\text{b}$ are evaluated at these positions for each of the $N_\text{s}$ excitations, resulting in a matrix $\bm{\mathsf{P}}_\text{b}$ of size $6 N_\text{d} \times N_\text{s}$.

\item \textbf{Step 4}: The dyadic Green tensors between the $N_\text{d}$ numerical dipoles and the $N_\text{l}$ learning points are computed for all components of electric and magnetic dipole moments and electric and magnetic fields. This results in a matrix $\bm{\mathsf{G}}_\text{b}$ of size $6 N_\text{l} \times 6 N_\text{d}$.

\item \textbf{Step 5}: The global polarizability matrix $\bm{\mathsf{A}}$, of size $6 N_\text{d} \times 6 N_\text{d}$, is finally determined by solving an inverse problem from the entire learning set ($N_\text{s}$ excitations and as many scattering solutions). This is the key step of the GPM method. The inverse problem is solved in two consecutive steps. The first one consists in finding the dipole moments at the $N_\text{d}$ numerical dipole positions that generate the scattered field at the $N_\text{l}$ learning points. This is strictly equivalent to the radiation problems mentioned above~\cite{beghou2009synthesis,chardon2013sampling}. Here, we thus search for a function that minimizes the Euclidean distance between the reproduced solution $\bm{\mathsf{G}}_\text{b} \bm{\mathsf{X}}$ and the reference (true) solution $\bm{\mathsf{P}}_\text{s}^\text{true}$, that is
\begin{equation}\label{eq:inverse-problem-1}
\mathcal{I}_1 = \min_{\bm{\mathsf{X}}} \left \| \bm{\mathsf{P}}_\text{s}^\text{true} -  \bm{\mathsf{G}}_\text{b} \bm{\mathsf{X}} \right \|_2,
\end{equation}
where $\left \| . \right \|_2$ denotes the Euclidean norm. The second step then consists in finding the polarizability matrix that converts the background field generated by the $N_\text{s}$ sources into the dipole moments at the $N_\text{d}$ numerical dipole positions. We thus solve
\begin{equation}\label{eq:inverse-problem-2}
\mathcal{I}_2 = \min_{\bm{\mathsf{A}}} \left \| \bm{\mathsf{X}}^\text{H} -  \bm{\mathsf{P}}_\text{b}^\text{H} \bm{\mathsf{A}}^\text{H} \right \|_2,
\end{equation}
where $\bm{\mathsf{X}}^\text{H}$ denotes conjugate transpose of $\bm{\mathsf{X}}$.

In the method, we aim at reducing the number of numerical dipoles to a minimum. In general, we are therefore in the situation where $N_\text{l} \gg N_\text{d}$ and $N_\text{s} > 6 N_\text{d}$, in which case the two inverse problems, described by Eqs.~(\ref{eq:inverse-problem-1}) and (\ref{eq:inverse-problem-2}), are overdetermined. In practice, Eqs.~(\ref{eq:inverse-problem-1}) and (\ref{eq:inverse-problem-2}) are solved by evaluating the pseudoinverses of $\bm{\mathsf{G}}_\text{b}$ and $\bm{\mathsf{P}}_\text{b}^\text{H}$, respectively. The pseudoinverses were computed here using a singular value decomposition (SVD). Our initial numerical tests have shown that for the specific example presented below, standard regularization methods (e.g., truncated SVD, Tikhonov regularization, etc.~\cite{vogel2002computational}) are not leading to significant improvements. Further studies will nevertheless be welcome.

The GPM method exploits a total of $36 N_\text{d}^2$ degrees of freedom (i.e. the size of $\bm{\mathsf{A}}$) to predict the scattered field. In comparison, if one would use \textit{physical} (local and coupled) electric and magnetic dipoles interacting via Green tensors, one would have only $6 N_\text{d}$ degrees of freedom to solve the problem. We therefore expect that, for the same number of dipoles, sets of numerical dipoles provide more accurate reconstructions than set of physical dipoles. This is a crucial point for the study of the scattering properties of large ensembles of disordered particles, while maintaining reasonable computational loads. 

\item \textbf{Step 6}: Once the polarizability matrix $\bm{\mathsf{A}}$ is obtained, the scattered field for an arbitrary background field can be computed from Eq.~(\ref{eq:scattering-numerical-dipoles}) as
\begin{equation}\label{eq:scattered-matrix}
\bm{\mathsf{P}}_\text{s} = \bm{\mathsf{G}}_\text{b} \bm{\mathsf{A}} \bm{\mathsf{P}}_\text{b}.
\end{equation}
\end{itemize}

\subsection{Global polarizability matrix}

\begin{figure}
   \centering
   \includegraphics[width=0.9\columnwidth]{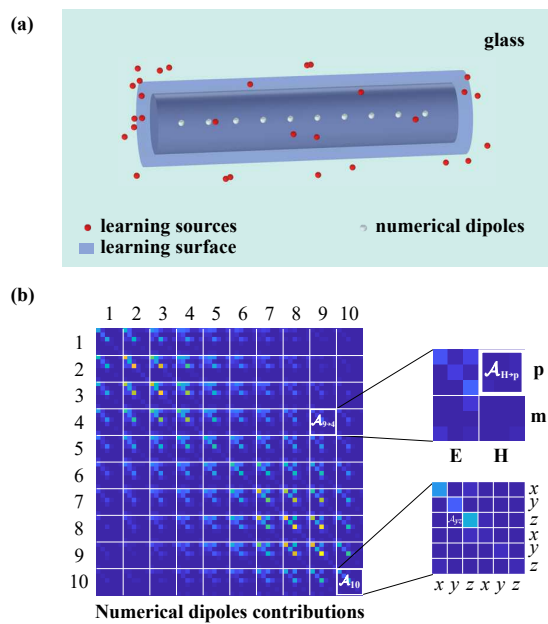}
   \caption{Global polarizability matrix. \textbf{(a)} We consider light scattering from a long dielectric cylinder. 30 localized sources (electric and magnetic point dipoles with polarizations along $x$, $y$ and $z$ directions) are randomly distributed at 40 nm from the particle boundary (red dots). These sources are successively used to excite the particle, resulting in $6 \times 30 = 180$ independent numerical fullwave simulations. The scattered field is evaluated at $N_\text{l}$ learning points on a surface lying at a distance $d_\text{l}$ (here $d_\text{l} = 20$ nm) from the particle boundary (semi-transparent blue surface). $N_\text{d}$ numerical dipoles (white dots, here $N_\text{d}=10$, regularly arranged on the cylinder axis) are used. \textbf{(b)} Polarizability matrix $\bm{\mathsf{A}}$, computed for $N_\text{d}=10$, $d_\text{l} = 20$ nm and $N_\textup{l}$ = 338. Each $6 \times 6$ block ($i$,$j$) describes the contribution of numerical dipole $i$ to the electric and magnetic moments induced in numerical dipole $j$. The insets show zooms of an off-diagonal block (here, $i=9$ and $j=4$) and an on-diagonal block (here, $i=j=10$, short notation used for simplicity).}
   \label{fig2}
\end{figure}

To illustrate the method, we consider the case of a high-index silicon cylinder with a large high-aspect-ratio, see Fig.~\ref{fig2}(a). The cylinder has a length $L=500$ nm and a diameter $d=100$ nm, and is excited at a wavelength $\lambda=2\pi/k_0$ of 580 nm, thereby leading to an efficient and multi-resonant light scattering (i.e., the cylinder exhibits several overlapping resonances at this wavelength). The Si permittivity is taken from tabulated data~\cite{green1995optical} ($\epsilon=15.8877 + 0.1796i$ at $\lambda= 580$ nm) and the ambient medium is glass ($\epsilon_\text{b}=2.25$). The rigorous fullwave simulations, required to acquire the large dataset (Step 1) and to validate our predictions with the numerical dipoles (Step 6), are performed with the commercial finite-elements software COMSOL Multiphysics~\cite{COMSOL}. Let us stress that the GPM method is not restricted to the use of a finite-elements software to build the learning set; any method rigorously solving Maxwell's equations may be used. Many light scattering softwares are in fact freely available nowadays~\cite{scattport}.

Simulations to build up the learning set are performed using electric and magnetic dipole sources oriented along the three Cartesian coordinates and placed at 30 different positions chosen randomly on a surface at a distance of 40 nm from the particle surface [red dots in Fig.~\ref{fig2}(a)], thereby amounting to $N_\text{s}=6 \times 30 = 180$ independent excitations. The choice of using localized sources in the near field of the particle instead of planewaves is motivated by our interest in simulating scattering by particles in close vicinity of each other. Numerical tests have shown that the use of localized sources is preferable to planewaves and that using more than 30 sources does not further improve the quality of the reconstruction. The learning points are placed regularly on a virtual surface [semi-transparent blue surface in Fig.~\ref{fig2}(a)] at a distance $d_\text{l}$ from the particle boundary. The results presented below are obtained with $N_\text{l} \approx 350$ learning points. We observed that further increasing $N_\text{l}$ does not improve the quality of the reconstruction. An appropriate set of $N_\text{d}$ numerical dipoles finally needs to be chosen. Here, we take the most natural arrangement that is a regular array placed on the cylinder axis. This turns out being a good compromise between computational load (driven in particular by $N_\text{d}$) and field reconstruction quality.

Figure~\ref{fig2}(b) shows the absolute value of each element of the global polarizability matrix $\bm{\mathsf{A}}$, computed with 10 numerical dipoles ($N_\text{d}=10$) for a distance between the learning surface and the particle boundary $d_\text{l}$ of only 20 nm. The computed matrix is approximately block-symmetric in this case because the numerical dipoles spatial distribution shares the symmetry properties of the cylinder and the learning sources were distributed randomly around it. This is however not a general property of the global polarizability matrix. Each $6 \times 6$ block describes how electric and magnetic fields induce electric and magnetic dipole moments (along the three Cartesian coordinates). The non-zero off-diagonal blocks describe the spatially non-local nature of $\bm{\mathsf{A}}$. As discussed above, these terms are additional degrees of freedom compared to systems of physical dipoles.

\subsection{Field reconstruction for known excitations}

\begin{figure}
   \centering
   \includegraphics[width=0.9\columnwidth]{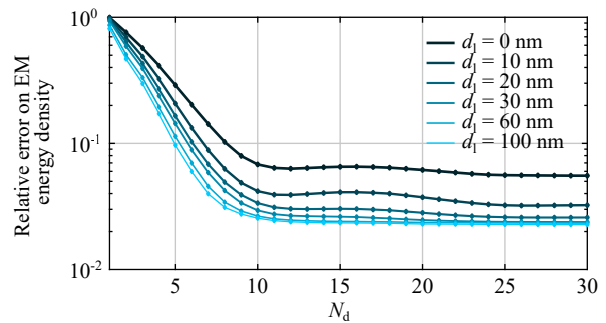}
   \caption{Convergence of the scattered fields on the learning surface. The relative error on the electromagnetic energy density computed with the numerical dipoles compared to the learning set is averaged over all points of the surface and all learning sources. A fast convergence of the error with the number $N_\textup{d}$ of numerical dipoles is obtained for the configuration of Fig.~\ref{fig2}. The accuracy of the reconstruction also improves when increasing the distance $d_\text{l}$ between the learning surface and the particle boundary.}
   \label{fig3}
\end{figure}

To get a first estimate on the quality of the field reconstruction following the numerical resolution of the inverse problem, we compare the scattered electromagnetic energy density $u=\frac{1}{2}(\mathbf{E}_\text{s} \cdot \mathbf{D}_\text{s} + \mathbf{B}_\text{s} \cdot \mathbf{H}_\text{s})$ on the virtual surface as predicted from the numerical dipoles using Eq.~(\ref{eq:scattered-matrix}) and obtained from the learning set. Figure~\ref{fig3} shows the relative error between the prediction and reference value, averaged over the learning surface and over all sources used for learning, as a function of the number $N_\textup{d}$ of numerical dipoles and for varying distances $d_\textup{l}$ between the learning surface and the particle boundary. A clear trend is observed. First, a minimal number of numerical dipoles is needed to get a plateau in the convergence curve. This is expected for such large particles, which do not radiate as a simple electric and/or magnetic dipole in the far field. Second, additional numerical dipoles beyond $N_\text{d} \approx 10$ do not lead to significant improvements of the reconstruction. Third, accurate predictions are more difficult to achieve when the learning surface approaches the particle boundary, because the field undergoes faster spatial variations.

\begin{figure*}
   \centering
   \includegraphics[width=0.9\textwidth]{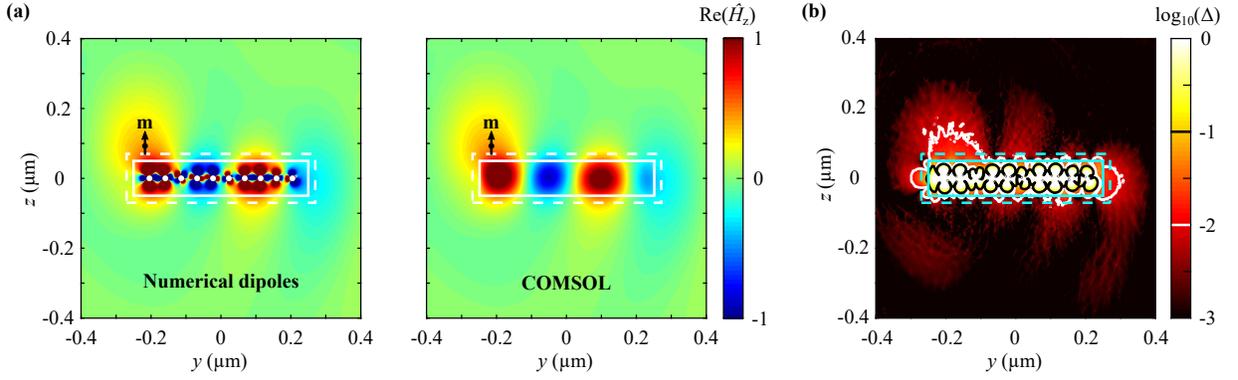}
   \caption{Reconstruction of the total field around the particle for one of the learning sources. The cylinder is illuminated by one of the 180 learned sources, a $z$-polarized magnetic point dipole at position $(x,y,z)=(-24,-219,90)$ nm. \textbf{(a)} $yz$-map of the $z$-component of the total magnetic field, normalized to its maximum beyond the learning surface. The cylinder surface is indicated by a solid white line and the learning surface by the dashed white line. 10 numerical dipoles (white dots) are regularly distributed on the cylinder axis. \textbf{(b)} Log of the difference between the two fieldmaps with 1\% and 10\% error contours. Errors superior to 10\% can only be observed in the particle volume. Errors superior to 1\% can be found near the source and at the ends of the cylinder but mainly stay confined within the volume bounded by the learning surface, thereby proving good agreement on the near field.}
   \label{fig4}
\end{figure*}

Next, we test how well the field is reconstructed everywhere in space for one of the learning sources by computing a near-field map around the particle using the numerical dipoles and comparing it with the reference data. In Fig.~\ref{fig4}(a), we show the real part of the total magnetic field ($z$-component) generated from one of the dipole sources used for learning, as predicted from the GPM method and computed with the finite-element method. The fields were normalized to the maximum field amplitude outside the particle. The divergence of the Green tensor naturally yields important discrepancies in the particle volume (bounded by a solid white line). These are not of our concern here. More importantly, the agreement between the fields near and beyond the learning surface (indicated by a dashed white line) is excellent by eye. Figure~\ref{fig4}(b) shows the point-by-point relative error between the two fieldmaps in $\log_{10}$-scale. Errors superior to 10\% are restricted to the particle volume and errors superior to 1\% hardly go past the learning surface, except near the source and at both ends of the cylinder. The maximum error beyond the learning surface is 3.9\%, proving a good agreement of the fields overall. The quality of the field reconstruction slightly varies with the considered source and the retrieved field components, but it overall remains very good, making us confident about the validity of the method.

\subsection{Generalization for arbitrary excitations}

\begin{figure}
   \centering
   \includegraphics[width=0.9\columnwidth]{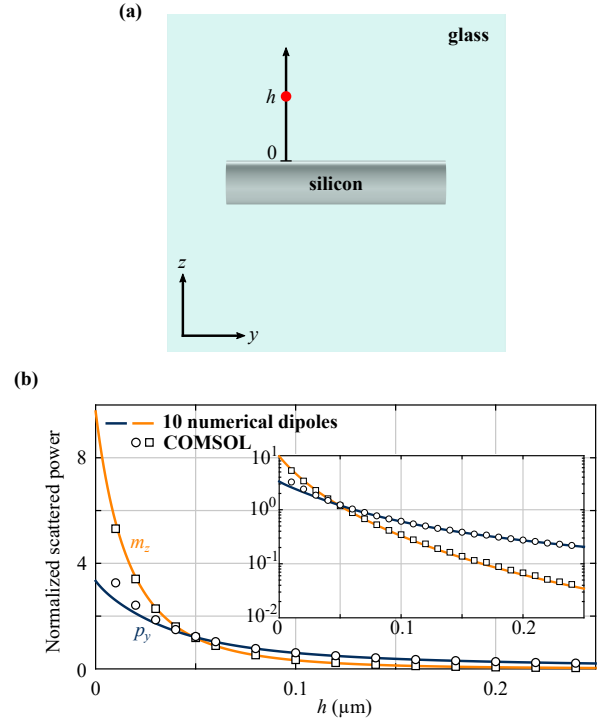}
   \caption{Generalization capability for unknown dipole sources. \textbf{(a)} A point dipole source is placed at varying heights $h$ above the particle surface (lateral position at $1/4$ of the cylinder length). \textbf{(b)} Scattered power normalized by the power emitted by the source in the uniform background, for varying distances $h$ and two different dipole sources $p_y$ and $m_z$, as calculated from the GPM method (solid lines) and with the reference (COMSOL, white markers). A quantitative agreement is achieved for sources positions $h \gtrsim 30$ nm.}
   \label{fig5}
\end{figure}

The GPM method has so far been validated for excitations that were used during learning to compute the polarizability matrix $\bm{\mathsf{A}}$. Our objective is however to predict the scattered field for an arbitrary excitation. To test this, we first consider a localized dipole source at various distances $h$ from the particle surface, see Fig.~\ref{fig5}(a) and calculate the power scattered by the particle. Figure~\ref{fig5}(b) shows a comparison between results obtained with the GPM method (solid lines) and the reference data (white markers) for a $y$-polarized electric dipole source $p_y$ and a $z$-polarized magnetic dipole source $m_z$. As sources get farther away from the particle, the scattered power naturally decreases. A quantitative agreement between predictions and reference data is observed for source positions $h \gtrsim 30$ nm. Deviations are observed when the sources approach the particle surface, especially for $p_y$. Similar trends are found for all other dipole sources. These deviations are expected since the learning surface is at $d_\text{l} = 20$ nm and the sources used for learning were placed at 40 nm from the particle boundary.

\begin{figure*}
   \centering
   \includegraphics[width=0.9\textwidth]{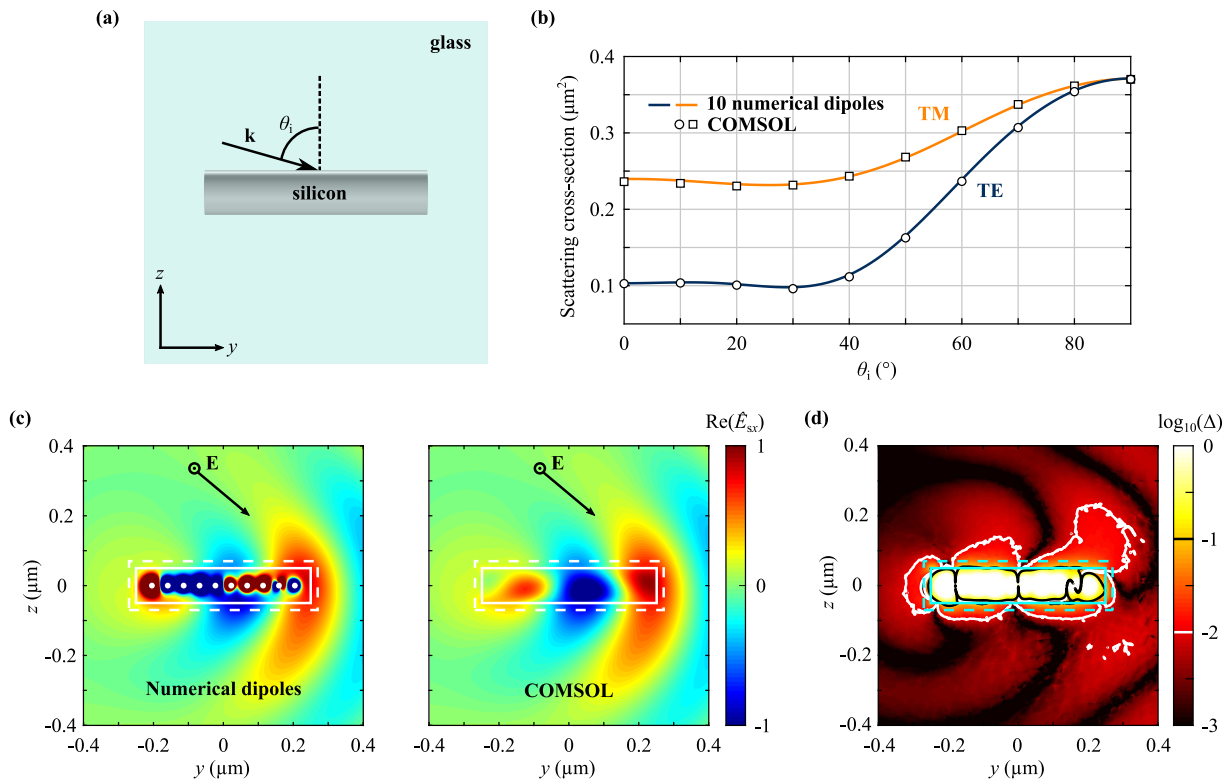}
   \caption{Generalization to planewave excitations. \textbf{(a)} The cylinder is illuminated by a planewave at an angle $\theta_\text{i}$. \textbf{(b)} Scattering cross-sections for TE and TM polarizations. Predictions from numerical dipoles (solid lines) are in very good agreement with reference data (COMSOL, white markers). \textbf{(c)} Maps of the scattered electric field ($x$-component) for TE-polarization at $\theta_\text{i}=50^\circ$, normalized to the maximum of the scattered field beyond the learning surface. The cylinder and numerical dipoles are the same as in the previous figures. \textbf{(d)} Difference between the field maps in $\log_{10}$-scale with 1\% and 10\% error contours.}
   \label{fig6}
\end{figure*}

Then, we consider a linearly-polarized planewave incident in the $yz$-plane at an angle $\theta_\text{i}$ defined from the normal to the cylinder axis along the $z$-direction [Fig.~\ref{fig6}(a)], and compute the scattering cross-section of the particle with varying incident angles. Results are shown in Fig.~\ref{fig6}(b) for both TE and TM polarizations. A quantitative agreement is again observed between our predictions and reference data, the maximum relative error being 1.6\% for the TE-planewave at $\theta_\text{i}=50^\circ$. In Figs.~\ref{fig6}(c)-(d), we show the fieldmaps of the scattered electric field (real part, $x$-component) as well as their difference in $\log_{10}$-scale. Except for the fields inside the cylinder, it is hard to notice any difference by eye beyond the learning surface. We remark that the relative error map reveals slightly larger errors around the specular reflection direction, which however remain quite acceptable (maximum of 4.5\%).

It is finally interesting to estimate the benefit brought by the (non-local) numerical dipoles set in the GPM method in comparison with the popular discrete dipole approximation (DDA), which relies on sets of local physical dipoles. For illustration, we consider the scattering properties of an individual cylinder illuminated by a planewave at $\theta_\text{i}=90^\circ$. Table~\ref{tab:performance} compares the computed scattering cross-sections, relative errors with respect to the reference data obtained with COMSOL, and the user time and physical memory required to perform the computation with the GPM method, implemented in MATLAB 2018a~\cite{MATLAB}, and the DDA using DDSCAT 7.3~\cite{ddscat,draine1994discrete}. The computations were launched on the same (standard) laptop. One observes that the DDA has difficulties reaching good accuracy on this scattering problem by a high-index resonant particle. An isolated example is not sufficient to draw general conclusions on the performance of a method -- a thorough benchmark study performed on various particles and various excitations would be needed --, nevertheless this gives a first evidence that the GPM method can indeed provide accurate results with significantly reduced computational load, when compared to a method based on physical dipoles.

\begin{table*}[htbp]
\centering
\caption{\bf Comparison of computational accuracy, user time and physical memory usage between the GPM method and a numerical method based on physical (coupled) dipoles for the scattering cross-section of the silicon cylinder upon planewave excitation at $\theta_\text{i}=90^\circ$.}
\begin{tabular}{ccccc}
\hline
Numerical method & GPM & \multicolumn{3}{c}{Discrete dipoles} \\
Software\footnote[1]{The GPM script on MATLAB 2018a~\cite{MATLAB} and DDSCAT 7.3~\cite{ddscat} were launched on the same laptop equipped with Intel(R) Core(TM) i5-6300 U CPU\symbol{64}2.40 GHz and 8 Gb of RAM. The default computation parameters were used for DDSCAT.} & MATLAB 2018a script & \multicolumn{3}{c}{DDSCAT 7.3} \\
Number of elements\footnote[2]{The number of elements corresponds to $2N_\text{d}$ for the GPM method (accounting for electric and magnetic dipoles) and to the number of physical dipoles, or domains, in DDSCAT.} & 20 & 2 080 & 16 640 & 107 400  \\
\hline
$\sigma_\text{s}$ ($\mu$m$^2$)\footnote[3]{Finite-elements calculations with COMSOL Multiphysics 5.4, using a very refined mesh with 154 941 elements, lead to a scattering cross-section of 0.3702 $\mu$m$^2$.} & 0.3741 & 0.2197 & 0.3334 & 0.3595 \\
Relative error ($\%$)\footnote[4]{The relative error is calculated with respect to the COMSOL value reported above.} & 1.05 & 40.65 & 9.94 & 2.89 \\
User time (s) & 0.271\footnote[5]{The reported user time for the GPM method does not include the time associated to fullwave simulations required to build the learning set, since this initial computational effort depends on the solver used and only needs to be performed once.} & 0.329 & 3.086 & 14.762 \\
Physical memory (Mb) & 1.91 & 38.4 & 56.0 & 168.3 \\
\hline
\end{tabular}
  \label{tab:performance}
\end{table*}

In sum, we have shown that the GPM method can reproduce and predict the field scattered by an elongated and resonant particle with high accuracy even with a low number of numerical dipoles. This study has been performed for a specific particle at a specific wavelength and with a specific learning set (near-field sources and learning surface). We believe that these first results are very encouraging and promising for the modelling of complex particles. In the next section, we will show that the GPM method also leads to reliable predictions for multiple scattering by ensembles of particles in stratified media.

\newpage

\section{Multiple scattering}
\label{multiplescattering}

\subsection{Formalism}

We consider the general case of an ensemble of $N_\text{p}$ particles in a stratified medium, where the $m$-th particle is described with $N_m$ numerical dipoles (previously denoted as $N_\text{d}$). We define the volume $\Omega = \cup_{m=1}^{N_\text{p}} \Omega_m $ as the union of the set of inner volumes $\Omega_m$ delimited by the learning surfaces of the particles. The particles may not be all identical and they can be placed in different layers with different refractive indices, thereby leading to different global polarizability tensors $\bm{\mathcal{A}}^m$.

Classically, the field scattered by the ensemble of particles is the sum of the contributions of the $N_\text{p}$ particles. In the context of numerical dipoles, we thus have
\begin{eqnarray}\label{eq:multiple-scattering-numerical-dipoles}
\bm{\Psi}_\text{s} (\mathbf{r}) &=& \sum_{m=1}^{N_\text{p}} \sum_{i,j=1}^{N_m} \bm{\mathcal{G}} (\mathbf{r},\mathbf{r}_{jm}) \bm{\mathcal{A}}^m(\mathbf{r}_{jm},\mathbf{r}_{im}) \bm{\Psi}_\text{exc}^{im} (\mathbf{r}_{im}), \nonumber \\
& & \text{for} \; \mathbf{r} \notin \Omega,
\end{eqnarray}
where $\mathbf{r}_{im}$ is the position of the $i$-th numerical dipole of the $m$-th particle, $\bm{\mathcal{G}}$ is the electromagnetic dyadic Green tensor describing electromagnetic propagation in the stratified medium~\cite{paulus2000accurate}, and $\bm{\Psi}_\text{exc}^{im}$ is the field exciting the $i$-th numerical dipole of the $m$-th particle. Note that, compared to Eq.~(\ref{eq:scattering-numerical-dipoles}) which describes the field scattered by a single particle in a uniform background, the numerical dipoles are excited by a field in the stratified medium and radiate in the stratified medium as well. Equation~(\ref{eq:multiple-scattering-numerical-dipoles}) formally describes the scattered field everywhere outside the learning surfaces of the individual particles ($\mathbf{r} \notin \Omega$).

\begin{figure}
   \centering
   \includegraphics[width=0.9\columnwidth]{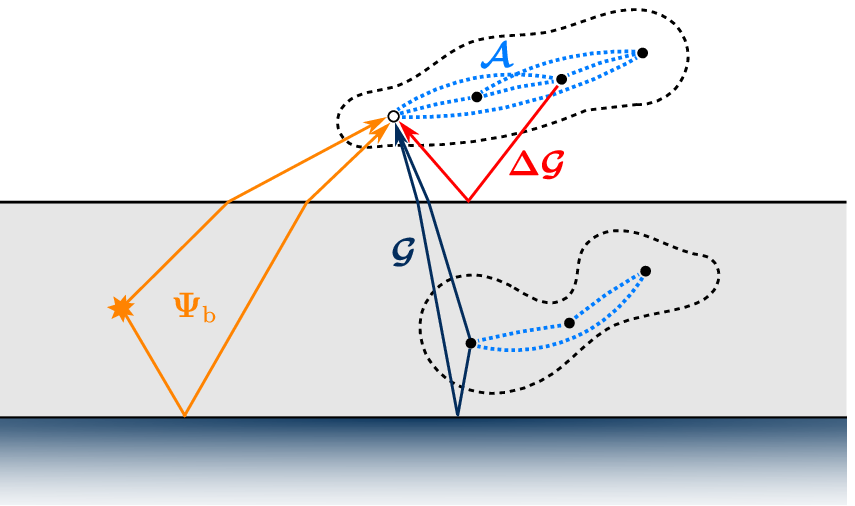}
   \caption{Implementation of multiple scattering in stratified media with the GPM method. In the general case, particles may have different shapes and be incorporated in different layers. The excitation field of a specific numerical dipole (white disk) is due to the background field in the stratified medium (orange arrows) and to the field radiated by the other numerical dipoles (black disks). Numerical dipoles belonging to different particles interact via the total Green tensor $\bm{\mathcal{G}}$ in the stratified medium (black arrows), while numerical dipoles belonging to the same particle interact only via the variation of the Green tensor $\bm{\Delta\mathcal{G}}$ in the stratified medium (red arrows). Some arrows have been removed for clarity.}
   \label{fig7}
\end{figure}

The first step in multiple scattering calculations consists in evaluating $\bm{\Psi}_\text{exc}$. For the $m$-th particle, this field is not only the background field but also the field scattered by all the other particles ($n \neq m$) in the stratified medium~\cite{mishchenko2006multiple}. Ensembles of physical dipoles, as in the discrete dipole approximation, are coupled to each other via the total Green tensor in the medium, independently of whether they belong to the same particle or not. This is not the case in the present method, since the polarizability tensor $\bm{\mathcal{A}}^m$ already considers the moments induced at all dipole positions by the exciting field within an individual particle. As illustrated in Fig.~\ref{fig7}, contrary to numerical dipoles belonging to different particles which can interact via the full Green tensor, those belonging to the same particle only interact via the interfaces of the stack, described by the variation of the Green tensor $\bm{\Delta\mathcal{G}}=\bm{\mathcal{G}} - \bm{\mathcal{G}}_\text{b}$. This is a direct consequence of the non-local nature of the polarizability tensor. We thus have
\begin{eqnarray}\label{eq:multiple-scattering-numerical-dipoles-exciting-field}
\bm{\Psi}_\text{exc}^{kn} (\mathbf{r}_{kn}) &=& \bm{\Psi}_\text{b} (\mathbf{r}_{kn}) + \sum_{m=1}^{N_\text{p}} \sum_{i,j=1}^{N_m} \Big[ \bm{\mathcal{G}} (\mathbf{r}_{kn},\mathbf{r}_{jm}) \nonumber \\
&-& \delta_{mn} \bm{\mathcal{G}}_{\text{b},m} (\mathbf{r}_{kn},\mathbf{r}_{jm}) \Big] \bm{\mathcal{A}}^m(\mathbf{r}_{jm},\mathbf{r}_{im}) \bm{\Psi}_\text{exc}^{im} (\mathbf{r}_{im}),
\end{eqnarray}
where $\delta_{mn}$ is the Kronecker delta ($\delta_{mn}=1$ for $m=n$ and 0 for $m\neq n$). Here, $\bm{\Psi}_\text{b}$ is the field created by a source in the stratified medium (orange arrows in Fig.~\ref{fig7}) and $\bm{\mathcal{G}}_{\text{b},m}$ is the Green tensor in the uniform background medium around the $m$-th particle. The numerical dipoles are by construction within $\Omega$. The sum on the right-hand side of Eq.~(\ref{eq:multiple-scattering-numerical-dipoles-exciting-field}) should not be interpreted as the field scattered by all particles other than the one of interest, as in the classical coupled dipole formalism. This quantity would instead take the form of Eq.~(\ref{eq:multiple-scattering-numerical-dipoles}).

Numerically, we first solve Eq.~(\ref{eq:multiple-scattering-numerical-dipoles-exciting-field}) via a standard matrix inversion to get the field exciting all numerical dipoles, and then solve Eq.~(\ref{eq:multiple-scattering-numerical-dipoles}) to get the scattered field at all points of space. The numerical resolution was performed with a home-made code running on MATLAB that has been successfully used in previous studies~\cite{langlais2014cooperative, jouanin2016designer, hugonin2017photovoltaics}.

\subsection{Multiple scattering in the near-field regime}

Let us now test the validity of the GPM method to solve multiple scattering problems, especially in cases of strong near-field interaction. We use the global polarizability matrix computed in Sec.~\ref{numericalmethod} with $N_m=10$ numerical dipoles.

\begin{figure*}
   \centering
   \includegraphics[width=0.9\textwidth]{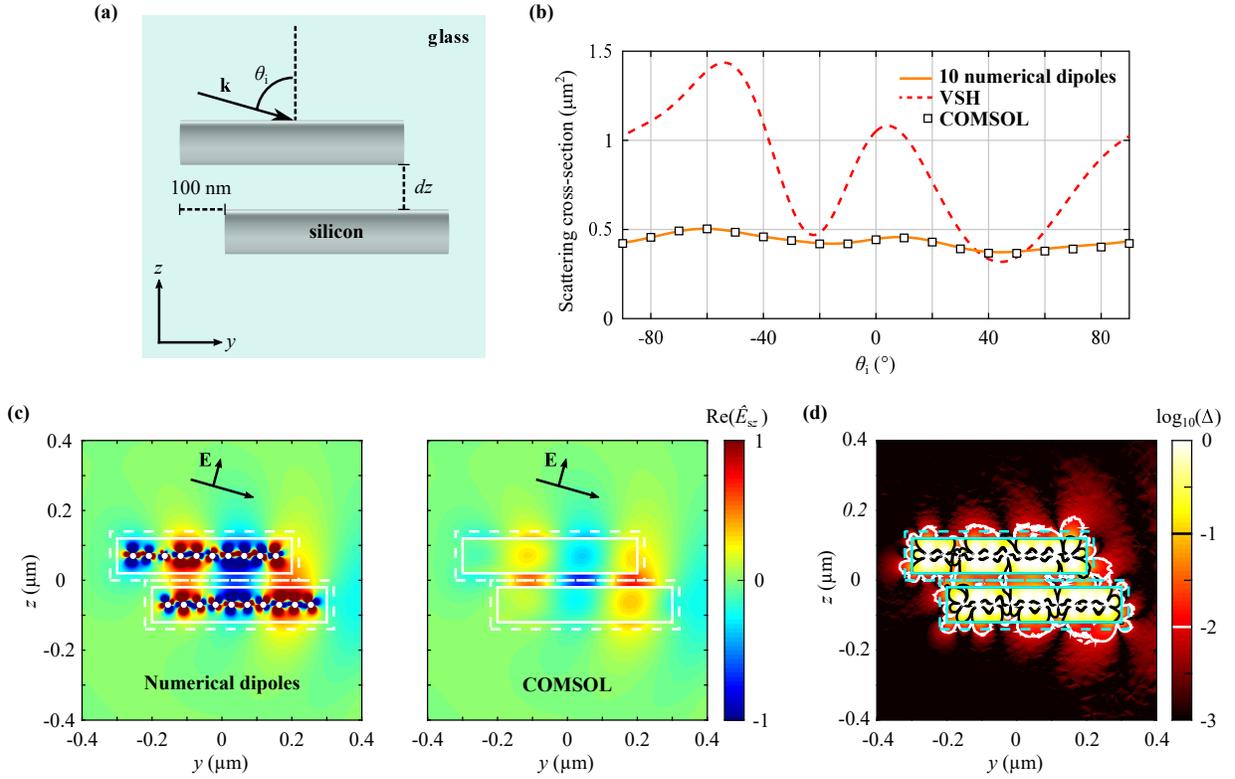}
   \caption{Modelling of two particles interacting in their near field. \textbf{(a)} Two parallel cylinders, shifted by 100 nm along $y$ and separated by a distance $dz$, are illuminated by a planewave with incident angle $\theta_\text{i}$. \textbf{(b)} Scattering cross-sections of the cylinder pair with $dz=40$ nm, for TM polarization with varying $\theta_\text{i}$, as predicted from the GPM method with numerical dipoles (solid line), the T-matrix method with VSH decomposition (dashed line) and COMSOL (white markers). While the T-matrix method expectedly fails to predict the response of the system for such strongly interacting particles, the GPM method is in very good agreement with exact simulations. \textbf{(c)} Scattered electric field maps, normalized to the maximum amplitude outside the cylinder, for a TM planewave at $\theta = 80^\circ$ for the system with $dz=40$ nm. \textbf{(d)} Difference between the field maps from (c) with 1\% and 10\% error contours, shown in $\log_{10}$-scale. Errors superior to 1\% hardly go past the learning surface proving the good near-field prediction.}
   \label{fig8}
\end{figure*}

We first consider two parallel cylinders shifted by 100 nm along the $y$-direction, as illustrated in Fig.~\ref{fig8}(a). The two cylinders are separated by a distance $dz$ and illuminated by a planewave with incident angle $\theta_\text{i}$. To start, we fix $dz = 40$ nm, which is the critical situation where the two learning surfaces are in contact. Figure~\ref{fig8}(b) shows the scattering cross-section of the particle pair for TM polarization at several angles of incidence, as predicted from the GPM method (solid line) and computed with COMSOL (white markers). For the sake of comparison, we also compute the predictions obtained with the T-matrix method relying on VSHs (dashed line) and thus limited in applicability to the circumscribing sphere. As expected in such situations of near-field interaction, the VSH representation of individual particles largely fails to predict the scattering response. In contrast, results from the GPM method are in quantitative agreement with the reference data, the maximum error being 4\% for $\theta = 80^\circ$. The near-field maps and their difference are shown in Figs.~\ref{fig8}(c)-(d). Let us note that even the field in the gap between the two cylinders is fairly well predicted.

\begin{figure}
   \centering
   \includegraphics[width=0.9\columnwidth]{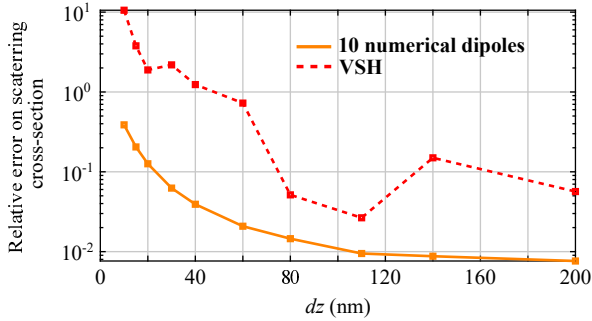}
   \caption{Relative error in the scattering cross-section computed for the configuration in Fig.~\ref{fig8} (TM polarization, $\theta_\text{i} = 80^\circ$) as a function of distance $dz$ between the two cylinders. The relative errors with respect to COMSOL results increase with decreasing distances $dz$. The GPM method with its numerical dipoles is better than the T-matrix method with VSH decomposition (here, up to the 11th multipole order) by an order of magnitude. Note that the T-matrix method is only expected to be valid at distances $dz \gtrsim 410$ nm.}
   \label{fig9}
\end{figure}

To gain insight on the limits of our approach, we study the evolution of the results in the worst case (TM polarization at $\theta = 80^\circ$) as a function of $dz$. In Fig.~\ref{fig9}, we show the relative error between the reference data and our predictions (solid line) and the T-matrix method predictions (dashed line). The agreement improves with increasing separation distance, with relative errors below 1\% for $dz \gtrsim 110$ nm, which is about the error made on individual particles, see Fig.~\ref{fig6}. Expectedly, the error increases significantly with stronger near-field interaction. The error made with the VSHs is systematically greater than that made with numerical dipoles by about an order of magnitude. Let us emphasize however that the latter depends on the maximum multipolar order taken for the VSH expansion. Nevertheless, this comparison clearly shows the benefit of our method to deal with particles in close vicinity.

\begin{figure*}
   \centering
   \includegraphics[width=0.9\textwidth]{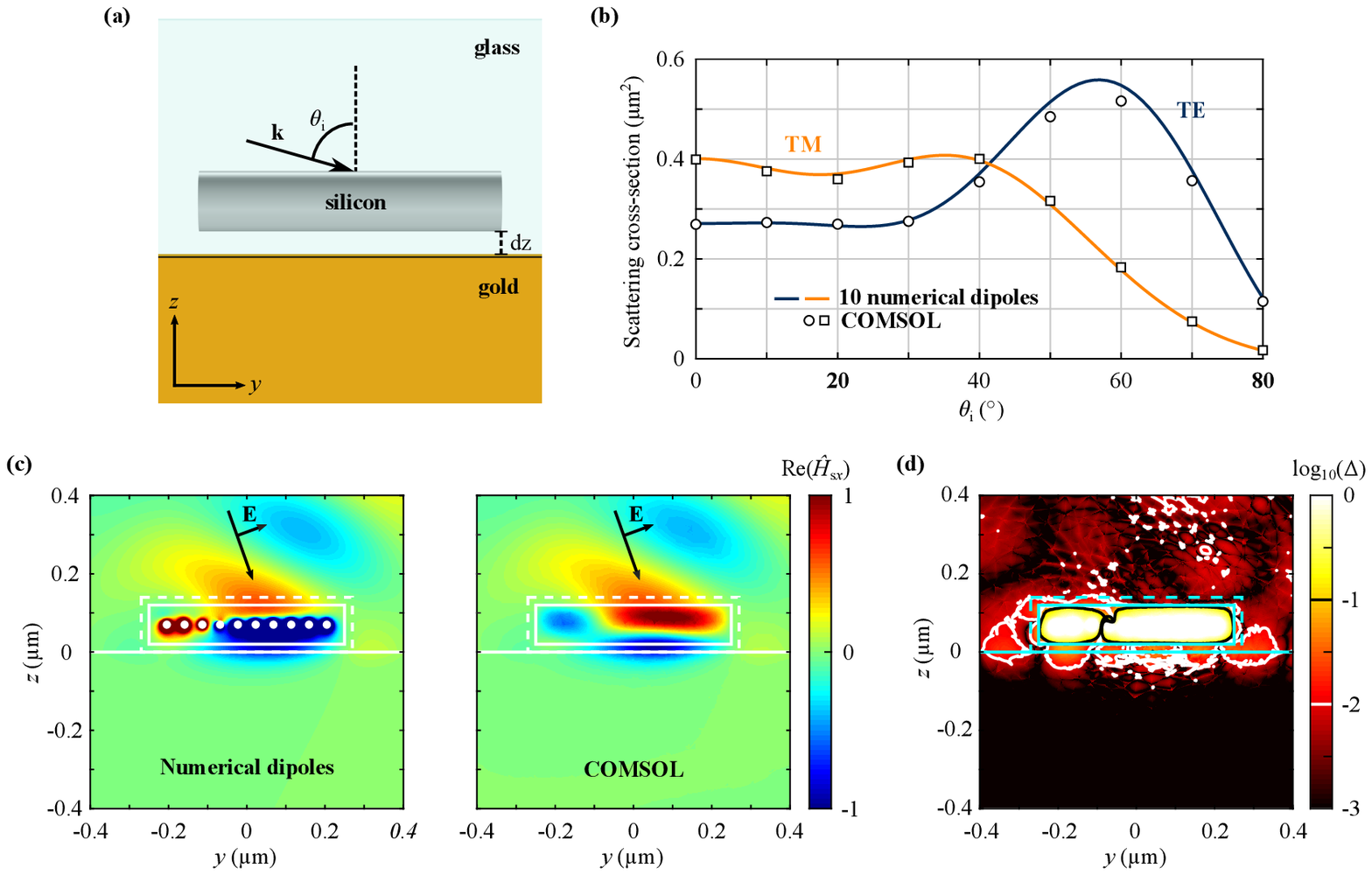}
   \caption{Modelling of one particle in the vicinity of a metallic interface. \textbf{(a)} A cylinder, placed at a distance $dz$ above an interface, is illuminated by a planewave with incident angle $\theta_\text{i}$. \textbf{(b)} Scattering cross-sections of the cylinder with $dz=20$ nm, for TE and TM polarizations with varying $\theta_\text{i}$, as predicted from the GPM method (solid line) and as computed with COMSOL (white markers). A good agreement is found. \textbf{(c)} Scattered electric field maps, normalized to the maximum amplitude outside the cylinder, for a TM planewave at $\theta = 20^\circ$. \textbf{(d)} Difference between the field maps from (c) with 1\% and 10\% error contours, shown in $\log_{10}$-scale. Errors superior to 1\% are found mostly near the interface and in the direction of specular reflection. Nevertheless, the results are very good considering the extreme case considered here.}
   \label{fig10}
\end{figure*}

Next, we perform a similar study for an individual particle in close vinicity of a planar interface. We consider the critical situation in which the learning surface touches the interface, that is for $dz=20$ nm, see Fig.~\ref{fig10}(a). The gold substrate has a permittivity $\epsilon_\text{Au} = -8.7494 + 1.5808i$ at $\lambda = 580$ nm~\cite{palik1998handbook} and the system is excited by an incident planewave. In Fig.~\ref{fig10}(b), we show the scattering cross-section of the particle for both TE and TM polarizations as a function of the incident angle $\theta_\text{i}$. A maximum relative error of 6.1\% is observed for TE-polarization and $\theta_\text{i}=50^\circ$. The near-field maps are computed instead for a TM-polarized planewave at $\theta_\text{i}=20^\circ$ (relative error of 3\% on the scattering cross-section) since one could expect an efficient excitation of surface-plasmon polaritons at the dielectric-metal interface for this polarization. As shown on Figs.~\ref{fig10}(c)-(d), a good agreement is also found. Note in particular that the field at the metal-dielectric interface is correctly predicted. The accuracy of the field prediction naturally improves when the particle is moved farther apart from the interface, as in Fig.~\ref{fig9}.

\begin{figure*}
   \centering
   \includegraphics[width=0.9\textwidth]{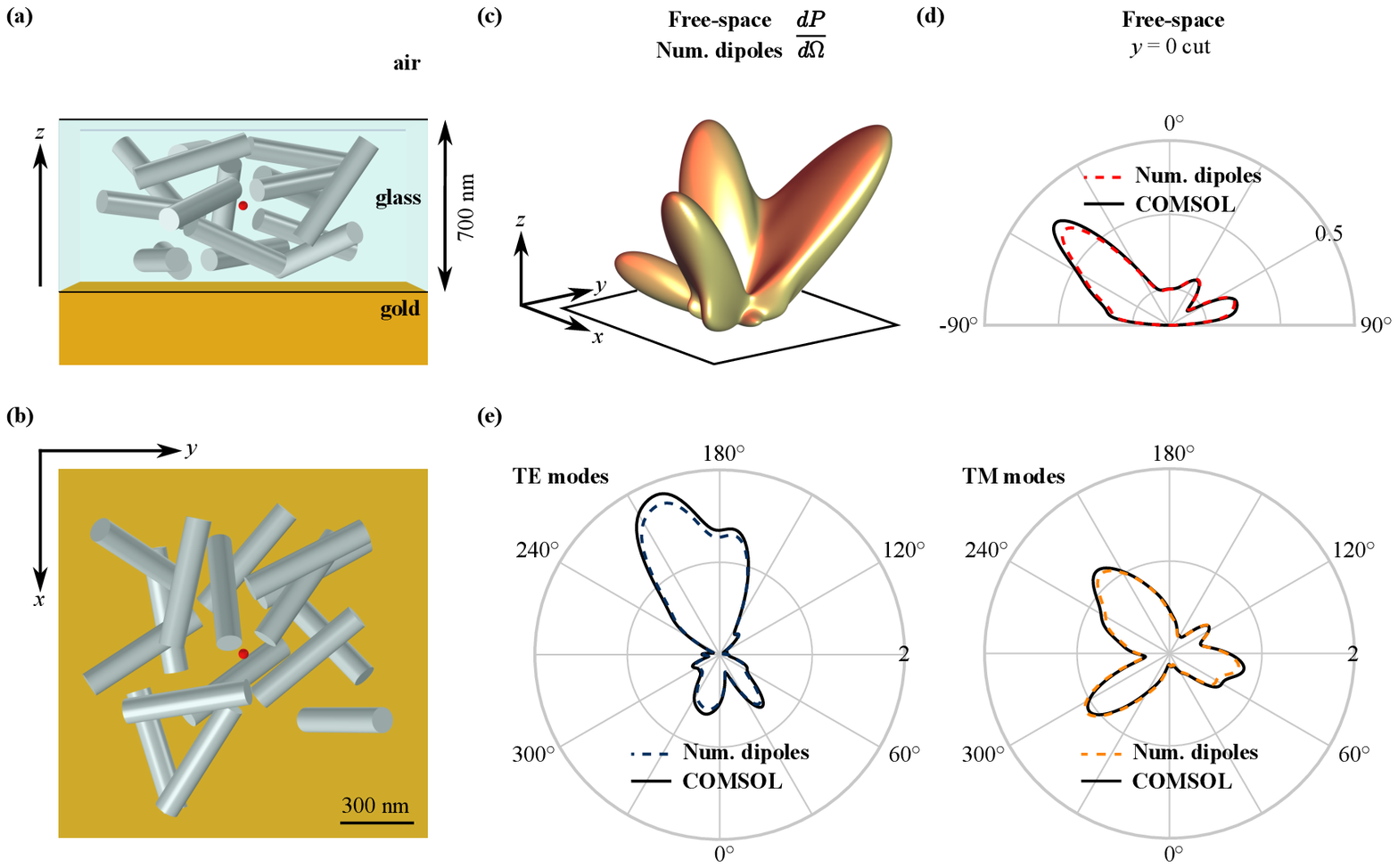}
   \caption{Modelling of a dense ensemble of particles in a stratified geometry. 16 silicon cylinders are placed randomly in a 700 nm thick glass layer deposited on a gold substrate. The system is excited by an $x$-polarized electric dipole placed in its center (red dot). \textbf{(a)} Lateral view ($xz$) of the system. \textbf{(b)} Top view ($xy$) of the system. \textbf{(c)} Free-space radiation diagram obtained with 10 numerical dipoles. \textbf{(d)} 2D cut of the free-space radiation diagram as computed with the GPM method (dashed line) and COMSOL (solid line). \textbf{(e)} Radiation diagrams in the guided modes, composed of 4 TE modes (left) and 4 TM modes (right), as computed with the GPM method and COMSOL. The agreement for both the free-space and the guided-mode radiation diagrams is very good considering the complexity of the present example. The COMSOL calculations required 78 Gb of physical memory to be completed, compared to 14.4 Mb for the GPM method implemented in MATLAB 2018a.}
   \label{fig11}
\end{figure*}

Finally, we test the method capability to simulate a dense ensemble of particles in optical stacks by considering a disordered ensemble of $N_\text{p}=16$ silicon cylinders embedded in a metallo-dielectric stack, see Fig.~\ref{fig11}(a)-(b). Compared to an analogous study made in a previous work~\cite{theobald2017plane}, the particles are larger in size and incorporated in a layered geometry. The array is also much denser than in a previous work made with an initial version of the method~\cite{jouanin2016designer}. We did not investigate larger systems due to the huge computational memory required to reach converged, trustable reference data. The stack consists of a 700-nm-thick glass layer on a gold substrate and an air cladding. The cylinder positions are determined by random sequential addition, imposing a minimum separation of 40 nm between all cylinders and between the cylinders and the two interfaces. To describe particles rotated with respect to their original frame, the polarizability matrix $\bm{\mathsf{A}}$ of each particle should be transformed using a rotation operator in Cartesian coordinates. The system is excited by an $x$-polarized electric dipole placed in the center of the system (red dot in Fig.\ref{fig11}(a)-(b)). The radiation diagrams $dP/d\Omega$ in free space and in the guided modes of the stack are computed with a freely available home-made software implementing a generalized near-to-far-field transformation (NFFT)~\cite{yang2016near}. They are obtained analytically from the field produced by the numerical dipoles for our method and from the field on the surface of a box surrounding the entire system for the finite-element calculations. Figure~\ref{fig11}(c) shows the radiation diagram in free space as computed with the GPM method. In Fig.~\ref{fig11}(d), we compare a cut of the free-space radiation diagrams ($xz$-plane at $y=0$) as predicted by our method and as computed with COMSOL and the NFFT software. Our method quantitatively reproduces the angular features of the radiation pattern, leading to a relative error near the diagram maximum of 7.3\% and 4.1\% on the scattered power (integral over the upper hemisphere). Similar conclusions are reached for the radiation into TE and TM guided modes (that involve photonic and plasmonic modes), where the relative errors are between 2 and 5\%.

All in all, these results are very encouraging considering the complexity of the investigated system and let us envision the possibility to simulate accurately large disordered systems composed of many particles with short and long-range interactions, that would be otherwise very difficult to simulate with brute-force Maxwell's equations solvers relying on full discretization.

\section{Conclusion}
\label{conclusion}

Light scattering by non-spherical resonant particles incorporated in optical stacks is a modelling challenge with potentially high impact in emerging photonic and plasmonic technologies. The difficulty lies in the multi-scale aspect of the problem. On the one hand, one should be able to model large ensembles of particles, interacting via free-space and guided modes on large distances compared to the wavelength. On the other hand, one should consider optical phenomena occurring at the scale of an individual particle, when interacting in the near-field with other particles and with interfaces.

In this article, we have introduced a new numerical method, named global polarizability matrix (GPM) method, that consists in finding a small set of electric and magnetic polarizable elements capable of reproducing the field scattered by a particle of arbitrary shape and composition for arbitrary near or far-field excitations, with high accuracy in the near and far-field regions, and at a low computational cost. This set of so-called ``numerical dipoles'' is described by a global and spatially non-local polarizability matrix providing more degrees of freedom for field reconstruction ($36 N_\text{d}^2$) compared to physical dipoles ($6 N_\text{d}$). The matrix is computed numerically by solving an inverse problem based on a learning set obtained from fullwave simulations for multiple independent excitations. These simulations can be performed with any Maxwell's equations solver. Simulations should be repeated for each particle parameter (refractive index and geometry), refractive index of the embedding medium and frequency, thereby requiring an important initial computational effort. However, once these initial computations are performed, the global polarizability matrix can be used to describe scattering by particles in arbitrary substrates and superstates, having arbitrary positions and orientations. Owing to a reduced number of elements to describe particle scattering and the existence of efficient numerical tools to compute Green tensors in stratified media, the method allows solving multiple scattering problems in large and dense ensembles of particles in layered geometries with strongly reduced computational load. Another advantage of this strategy is its flexibility, since the surface beyond which the field should be well predicted is fully determined by the user. It can be adapted depending on the problem of interest.

We have validated the method with the example of a large high-aspect-ratio high-index dielectric cylinder. First, we have shown that light scattering by the particle can indeed be well predicted for arbitrary sources once the polarizability matrix is computed, including when the excitation and the observation points are in the near-field region. Second, we have validated that the method could be used to predict scattering by dense ensembles of particles embedded in stratified media. We have systematically tested the validity of the approach in critical situations, which typically resulted in maximal errors of the order of a few percents.

Several open questions are left for future works. A first point concerns the optimization of the method performance. In this work, we have rapidly fixed a certain number of parameters to ease understanding. In particular, the learning set was obtained with localized sources placed in the near field of the particle at a fixed distance from its surface. Near-field sources display larger field variations, exciting higher-order particle resonances, which play a decisive role in near-field interaction problems. One could however use a broader range of excitations (localized sources at various distances from the particle, planewave sources) as a complement. It will be important to perform a systematic investigation of the role of the various input parameters on the predictive capability of the method. A related question, that is recurrent in numerical methods based on elementary sources and inverse problems, concerns the optimal spatial arrangement of the numerical dipoles in the volume of the particle. Here, we have placed the dipoles regularly on the cylinder axis, following the symmetry of the problem. This evidently has some limitations since it cannot represent high-order derivatives in directions normal to the axis. As we are attempting to minimize the number of numerical dipoles, it is so far unclear what the optimal arrangement would be. This study may be done following our physical intuition on optical resonances, but one may consider using more advanced tools~\cite{beghou2009synthesis,koyama2018joint} for this numerical optimization problem. Exploring many configurations of numerical dipoles is in fact not an issue, since, given a learning set, the computation of the polarizability matrix is very rapid.

Much remains to be understood also on the benefit of the GPM method in terms of computational performance and predictive capability compared to methods based on physical dipoles. In Table~\ref{tab:performance}, we have provided a comparison of the method with the discrete dipole approximation (DDA) on a specific example. The GPM method was shown to predict the scattering cross-section of the cylinder more accurately and more efficiently than DDA with a drastically lower computational load. This provides a first evidence of the benefit offered by numerical dipoles over physical dipoles. We further note that, to fully address the role of spatial non-locality on predictive capabilities, it would be preferable to compare the GPM method with an equivalent method using physical dipoles interacting via electromagnetic wave propagation with polarizabilities determined by a numerical inversion scheme on the same learning dataset as for the GPM method. Comparisons could be made either for the same polarizability matrix size or for the same number of degrees of freedom (meaning larger matrix sizes for physical dipoles).

Another point that should deserve attention concerns the generalization to other particles. This possibility is readily at hand thanks to the great variety and flexibility of rigorous numerical methods, such as the finite-elements method. Here, we have focused on a specific dielectric particle at a fixed frequency. It will be interesting in future studies to test the method on metallic or metallo-dielectric particles, possibly exhibiting sharp edges inducing hot spots~\cite{chung2011plasmonic}. Particles with complex shapes would also constitute an interesting test case for the method. The possibility to define a learning surface at will raises the question of whether one could predict the field between the two parallel arms of a U-shaped particle, for instance. The validation of the method may be done on finite, disordered ensembles of particles, as in this article, but also on periodic systems, since many reliable modelling tools are available.

Finally, since the final application of the GPM method is to study large ensembles of particles, it will be critical to study the convergence of the method with increasing number of particles. As we have seen in this work, errors of the order of a percent can be observed on an individual particle. This error may accumulate when increasing significantly the number of particles in strong near-field interaction, eventually leading to wrong predictions. Our first simulations on a set of 16 particles show errors on the order of a few percents, which remains quite reasonable. This result constitutes a first step towards the use of the approach modelling method in practical situations that will hopefully stimulate further efforts.

\section{Funding Information}

CNRS Mission for Interdisciplinarity (Project NanoCG), Programme IdEx Bordeaux – LAPHIA (ANR-10-IDEX-03-02, project X-STACKS), French National Agency for Research (ANR) (ANR-16-CE30-0008, project NanoMiX).

The authors thank Etienne Hartz and Nicolas Mielec for their initial contribution to the project.

\appendix

\section{Relation with the transition operator of a particle}
\label{App:Tmatrix}

In this Appendix, we wish to emphasize some similarities but also important differences between the transition operator $T$ of a particle and the global polarizability tensor introduced in this work. The derivation given here is completely classical (see, e.g., Ref.~\cite{mishchenko2006multiple}). We consider a particle with relative permittivity $\bm{\epsilon} (\mathbf{r}) = \epsilon_\text{b} \mathbf{I} + \bm{\Delta\epsilon}(\mathbf{r})$, where $\mathbf{r}$ is the position in Cartesian coordinates, $\epsilon_\text{b}$ is the permittivity of the background and $\bm{\Delta\epsilon}$ is the permittivity variation due to the particle. From Maxwell's equations, one straightforwardly shows that the field scattered by the particle, $\mathbf{E}_\text{s} = \mathbf{E} - \mathbf{E}_\text{b}$, where $\mathbf{E}$ is the total field and $\mathbf{E}_\text{b}$ the background field, satisfies a vector wave propagation equation
\begin{equation}\label{eq:scattered-wave-equation}
\bm{\nabla} \times \bm{\nabla} \times \mathbf{E}_\text{s} (\mathbf{r}) - k_0^2 \epsilon_\text{b} \mathbf{E}_\text{s} (\mathbf{r}) = k_0^2 \bm{\Delta\epsilon}(\mathbf{r}) \mathbf{E}(\mathbf{r}).
\end{equation}
The polarization field induced in the particle, $\mathbf{P}(\mathbf{r})/\epsilon_0 = \bm{\Delta\epsilon}(\mathbf{r}) \mathbf{E}(\mathbf{r})$, readily appears as a source for the scattered field in the background medium. The solution of Eq.~(\ref{eq:scattered-wave-equation}) can be expressed via the dyadic Green tensor in the background medium $\mathbf{G}_\text{b}(\mathbf{r},\mathbf{r}')$ as
\begin{equation}\label{eq:Green-function-total-field}
\mathbf{E}_\text{s} (\mathbf{r}) = k_0^2 \int \mathbf{G}_\text{b} (\mathbf{r},\mathbf{r}') \bm{\Delta\epsilon} (\mathbf{r}') \mathbf{E}(\mathbf{r}') d\mathbf{r}',
\end{equation}
where the integral is made over the particle volume due to the compactness of $\bm{\Delta\epsilon}$. We then rewrite Eq.~(\ref{eq:Green-function-total-field}) for the total field $\mathbf{E}$ as
\begin{equation}\label{eq:total-field}
\mathbf{E} (\mathbf{r}) = \mathbf{E}_\text{b} (\mathbf{r}) + k_0^2 \int \mathbf{G}_\text{b} (\mathbf{r},\mathbf{r}') \bm{\Delta\epsilon} (\mathbf{r}') \mathbf{E}(\mathbf{r}') d\mathbf{r}'.
\end{equation}
There exist various ways to solve this equation. Here, we focus on the Born expansion because it is very intuitive and analogous in some (but not all) aspects to the GPM method. Equation~(\ref{eq:total-field}) is solved iteratively by expressing the total field $\mathbf{E}$ in the integral on the right-hand side as a function of the background field $\mathbf{E}_\text{b}$, that is
\begin{eqnarray}
\mathbf{E} (\mathbf{r}) &=& \mathbf{E}_\text{b} (\mathbf{r}) + k_0^2 \int \mathbf{G}_\text{b} (\mathbf{r},\mathbf{r}') \bm{\Delta\epsilon} (\mathbf{r}')  \mathbf{E}_\text{b}(\mathbf{r}') d\mathbf{r}' \nonumber \\
&+& k_0^4 \int \mathbf{G}_\text{b} (\mathbf{r},\mathbf{r}') \bm{\Delta\epsilon} (\mathbf{r}')  \mathbf{G}_\text{b} (\mathbf{r}',\mathbf{r}'') \bm{\Delta\epsilon} (\mathbf{r}'') \mathbf{E}_\text{b}(\mathbf{r}'') d\mathbf{r}' d\mathbf{r}'' \nonumber \\
&+& ...
\end{eqnarray}
The first integral term describes the contribution of single scattering within the particle on the scattered field, the second integral term the contribution of double scattering within the particle, and so on. This can be conveniently expressed by defining the transition operator $\mathbf{T}$ of the particle, as
\begin{equation}\label{eq:operator-t}
\mathbf{T}(\mathbf{r},\mathbf{r}') = k_0^2 \; \bm{\Delta\epsilon} \left[ \delta(\mathbf{r}-\mathbf{r}') \mathbf{I} + \int \mathbf{G}_\text{b} (\mathbf{r},\mathbf{r}'') \mathbf{T}(\mathbf{r}'',\mathbf{r}') d\mathbf{r}'' \right],
\end{equation}
leading to
\begin{equation}\label{eq:operator-t-scattered-field}
\mathbf{E}_\text{s} (\mathbf{r}) = \int \mathbf{G}_\text{b} (\mathbf{r},\mathbf{r}') \mathbf{T}(\mathbf{r}',\mathbf{r}'') \mathbf{E}_\text{b}(\mathbf{r}'') d\mathbf{r}' d\mathbf{r}''.
\end{equation}

Equation~(\ref{eq:operator-t-scattered-field}) is evidently similar to Eq.~(\ref{eq:scattering-numerical-dipoles}) in that it relates a background field to a scattered field via a transition operator describing a radiating polarization density in the volume of the particle. As a matter of fact, generalizing Eq.~(\ref{eq:operator-t-scattered-field}) to magnetic fields and discretizing the electromagnetic transition operator as $\bm{\mathcal{T}}(\mathbf{r},\mathbf{r}') = \sum_{i,j=1}^{N_d} \delta(\mathbf{r}-\mathbf{r}_j) \bm{\mathcal{A}}(\mathbf{r},\mathbf{r}') \delta(\mathbf{r}'-\mathbf{r}_i)$ would directly lead to Eq.~(\ref{eq:scattering-numerical-dipoles}). Similarly to $\bm{\mathcal{A}}$, $\mathbf{T}$ is also intrinsic to the particle (at fixed $\omega$) and exhibits spatial non-locality. This spatial non-locality is related to multiple scattering within the volume of the particle and can only be neglected for tiny particles. Keeping the lowest order only in Eq.~(\ref{eq:operator-t}) is the so-called Born approximation, for which $\mathbf{T}$ becomes the particle polarizability.

The analogy between the classical transition operator and the global polarizability operator stops here. Indeed, as evidenced by Eq.~(\ref{eq:operator-t}), the operator $\mathbf{T}$ is a direct representation of the \textit{physical} (multiple-scattering) problem -- it depends on the local permittivity variation $\bm{\Delta\epsilon}$ and the spatial non-locality is driven by the Green tensor $\mathbf{G}_\text{b}$ -- whereas the tensor $\bm{\mathcal{A}}$ is a mathematical object that is not related explicitly to the permittivity variation (e.g., the numerical dipoles may be placed in regions where $\bm{\Delta\epsilon}=0$) and to the internal electromagnetic wave propagation.


\begin{thebibliography}{66}
\bibitem[{sca()}]{scattport}
\bibinfo{note}{See, for instance, the website scattport.org, which provides an
  updated list of freely available light scattering softwares.}
\end{thebibliography}

\end{document}